\documentclass[
  journal=pasa,
  manuscript=research-paper, 
  year=2020,
  volume=37,
]{cup-journal}

\usepackage{hyperref}

\usepackage{microtype,siunitx,booktabs}
\usepackage{tikz}

\usepackage{lineno}

\def\arcsec{\hbox{\hbox{$^{\prime\prime}$}}}

\usepackage{soul}

\newcommand\ergs{\ifmmode {\rm~erg\ s}^{-1} \else  ~erg s$^{-1}$\fi}
\newcommand\Zsun{\ifmmode Z_{\odot} \else $M_{\odot}$\fi}

\title{The radio source in Abell 980: A Detached-Double-Double Radio Galaxy?}

\author{Gopal-Krishna}
\affiliation{UM-DAE Centre for Excellence in Basic Sciences, University of Mumbai, Vidyanagari, Mumbai-400098, India}

\author{Surajit Paul}
\affiliation{Department of Physics, Savitribai Phule Pune University, Pune 411007, India}\email[Surajit Paul]{surajit@physics.unipune.ac.in}

\author{Sameer Salunkhe}
\affiliation{Department of Physics, Savitribai Phule Pune University, Pune 411007, India}

\author{Satish Sonkamble}
\affiliation{INAF-Padova Astronomical Observatory, Vicolo dell’Osservatorio 5, I-35122 Padova, Italy}

\doi{10.1017/pasa.2020.32}

\received {dd Mmm YYYY}
\revised  {dd Mmm YYYY}
\accepted {dd Mmm YYYY}
\published{22 September 2020}


\begin{document}

\begin{abstract}
It is argued that the new morphological and spectral information gleaned from the recently published LoFAR Two meter Sky Survey data release 2 (LoTSS-2 at 144 MHz) observations of the cluster Abell 980 (A980), in combination with its existing GMRT and VLA observations at higher frequencies, provide the much-needed evidence to strengthen the proposal that the cluster's radio emission comes mainly from two double radio sources, both produced by the brightest cluster galaxy (BCG) in two major episodes of jet activity. The two radio lobes left from the previous activity have become diffuse and developed an ultra-steep radio spectrum while rising buoyantly through the confining hot intra-cluster medium (ICM) and, concomitantly, the host galaxy has drifted to the cluster centre and entered a new active phase manifested by a coinciding younger double radio source. The new observational results and arguments presented here bolster the case that the old and young double radio sources in A980 conjointly represent a `double-double' radio galaxy whose two lobe-pairs have lost colinearity due to the (lateral) drift of their parent galaxy, making this system by far the most plausible case of a `Detached-Double-Double Radio Galaxy' (dDDRG).
\end{abstract}

\section{Introduction}\label{intro}
It is well known for several decades that the double radio sources hosted by early-type galaxies located inside galaxy clusters follow a significantly different evolutionary track as compared to those located outside the cluster environment \citep{McNamara_2007ARA&A,Fabian_2012ARA&A,Hardcastle_2013MNRAS,Pasini_2021MNRAS}. The difference arises mainly due to the higher ambient density, temperature (and the resulting static pressure as well as sound speed) encountered  by the relativistic plasma jets and the radio lobes inflated by them within the hot intra-cluster medium (ICM). Such interactions quicken the deceleration of the jets and weaken the associated Mach disks, leading to morphological transformation/deformation of the radio lobes. Secondly, a more efficient confinement of the lobes by the ICM static pressure delays their fading due to adiabatic expansion, thus allowing the synchrotron losses to accumulate and manifest as an ultra-steep spectrum (USS) of the radio  lobes \citep{1973MNRAS_baldwin}. Consequently, the USS radio lobes can be detected at low radio frequencies long after the cessation of energy injection by the jets, whereafter the lobes can rise within the ICM due to its buoyancy pressure \citep{Gull_1973Natur}. Such faded lobes are sometimes identified as X-ray ``cavities” \citep{bohringer_1993MNRAS, McNamara_2005Natur, birzan_2004ApJ}, or even become detectable as rejuvenated radio lobes (“Phoenix”) following their adiabatic compression by cluster merger shocks \citep{Ensslin_2001A&A}. Furthermore, as pointed out in many studies, radio sources can significantly contribute to moderating or even preventing the cooling flows \citep{fabian_1994ARA&A, birzan_2004ApJ}. Another dynamical effect, highlighted recently, is the bending of jets due to tension of the compressed ICM magnetic field \citep{chibueze_2021Natur}. The plethora of physical processes ongoing in galaxy clusters and the resulting rich phenomenology make them extremely interesting laboratories for studying the evolution of these largest gravitationally-bound structures in the universe. In this paper, we discuss new multi-band observations of a semi-relaxed cool-core cluster Abell 980 (A980), with particular attention to the ultra-steep-spectrum radio source complex embedded within the unusually hot ICM of this cluster, which display multiple indicators of episodic nuclear activity in its brightest cluster galaxy (BCG).

The BCG of the cluster A980 (RXC J1022.5+5006, $z$ = 0.1582;  \citealt{Ebeling_MNRAS_1996}) is known to possess a stellar halo of size $\sim80$~kpc, roughly co-spatial with the X-ray emission peak (see Fig.~\ref{fig:LoTSS_GMRT_optical_xray}). The cluster has an SZ mass of $M_{500}^{SZ} = 4.73_{-0.32}^{+0.29} \times 10^{14} \rm{M_{\odot}}$ \citep{planck2016A&A} and a relatively high X-ray luminosity ($L_X=7.1\times10^{44}$~erg~s$^{-1}$) originating from the ICM of size $\sim500$~kpc and a mean temperature T $\sim7.1$~keV, \citealt{Ebeling_MNRAS_1996}). Its WENSS image at 330 MHz (FWHM = 54{\arcsec}) gave the first hint of diffuse emission underlying a strong peak \citep{2009ApJRudnick}. From the radio observations with the GMRT (150/325 MHz) and VLA (1.5 GHz), the radio emission has been found to arise mainly from 3 components (A, B \& C, Fig.\ref{fig:LoTSS_GMRT_optical_xray}a), of which A and B are diffuse, with ultra-steep radio spectra (USS, $\alpha<-2$) and extending to the ICM outskirts, whereas the source C coinciding with the BCG is resolved into a smaller double radio source of size $\sim$ 55 kpc (\citealt{Salunkhe_2022A&A}; hereafter, S22). Here we present new observation-based evidence and arguments, particularly those emerging from the recently released high-sensitivity image with a 6 arcsecond resolution at 144 MHz (LoTSS/DR2, \citealt{Shimwell_2022A&A}), in order to reinforce the hypothesis associating all the 3 radio components (A, B, C) with the intermittent nuclear activity occurring in the BCG (S22). This 144 MHz image, when combined with the images obtained from the GMRT (325 MHz) and VLA (1.5 GHz) archival data, has enabled a clear delineation of the ultra-steep-spectrum (USS) components A \& B and defining their structural and spectral details.

A $\Lambda$CDM cosmology is assumed with parameters $H_{0}=70$ ${\rm{kms^{-1} Mpc^{-1}}}$, $\Omega_{0,m} =0.3$, $\Omega_{0,\Lambda}=0.7$ \citep{Condon_2018PASP} throughout this paper. The physical scale for the images is 2.73~kpc/arcsec.

\section{The data and analysis}\label{sec:data}

\subsection{The radio data}\label{sec:radio_anal}

A total of 314 minutes of on-source 325 MHz GMRT archival data (Project: ddtB020~\&~17\_073) and 56 minutes of  EVLA L-band (1-2~GHz, B-array) archival data (Project: 15A-270 and 12A-019) were processed for this study. Data from RR and LL correlations of GMRT 325~MHz were reduced and imaged using the Source Peeling and Atmospheric Modeling (SPAM) pipeline (for details, see \citealt{2017Intema_SPAM}). The EVLA L-band data from RR and LL correlations were analysed using the CASA calibration pipeline and images were produced by performing several rounds of phase-only self-calibration and one amplitude and phase self-calibration. The images were made using the various weighting and uvtaper parameters listed in Table~\ref{tab:imaging}. The 144~MHz high-resolution ($6^{''}\times6^{''}$) map of LoFAR Two meter Sky Survey data release 2 (LoTSS-2) has been taken directly from the LoTSS-2 archive \citep{Shimwell_2022A&A}.

\begin{table} 
\caption{Parameters of the radio images} %
\label{tab:imaging}      
\centering          
\begin{tabular}{lccccc}     
\hline\hline       
Frequency & Weighting & Robust & uv taper & FWHM & rms \\
(MHz)& Scheme & &  & (in \arcsec; PA in $^\circ$)& ($\mu$Jy/beam)\\
\hline  
325 & uniform$^w$ & -- & 5\arcsec* & $7.7\times6.7$; 169$^h$ & 150\\  
(GMRT)  & briggs$^s$ &  -2 &  -- & $8.2\times7.3$; 7$^l$ & 100\\
1500  & briggs$^c$ & -2 & -- & $2.8\times2.5$; 63$^h$ & 30\\
(EVLA)  & briggs$^c$ & 0 & 10 k$\lambda$ & $9.0\times8.7$; 31$^l$ & 25\\
\hline
\hline
\end{tabular}
\begin{tablenotes}
      \small
\item \textbf{Note}: Resolutions: $h$ - high, $l$ - low; images are produced in $w$ - WSCLEAN, $s$ - SPAM and $c$ - CASA. `*' In WSClean, 5 arcsec corresponds to $\sim 50$ $k\lambda$ for this image.
    \end{tablenotes}
\end{table}

The LOTSS-2 144~MHz and GMRT 325~MHz maps have been combined to produce the spectral index map. The SPAM calibrated GMRT visibilities at 325 MHz were taken to the WCSCLEAN algorithm \citep{Offringa_2014MNRAS}, to produce a 325 MHz image with the resolution of the LOTSS 144 MHz high-resolution map. Uniform weighting with a Gaussian taper of {5\arcsec} was used in WCSCLEAN. The small difference in the beam sizes of the two maps was then corrected by smoothing both maps to a common beam size of $8\arcsec\times8\arcsec$ (circular), in the CASA task IMSMOOTH. The images were re-gridded to the same pixel grid using CASA IMREGRID task before smoothing.  A minor astrometric offset between the LoTSS and GMRT images was corrected at the outset. Finally, the masking for these images was taken at $3\sigma$ isophotes of 325~MHz map. 

The spectral index was computed using the IMMATH task of CASA using the relation 
\begin{equation}\label{eq:Sp-Ind}
\alpha = \frac{\log(S_{\nu_2}/S_{\nu_1})}{\log(\nu_2/\nu_1)}
\end{equation}
where $S_{\nu_\#}$ and $\nu_\#$ are the values of flux density and the frequency of observation, respectively.

The spectral index error has been calculated with the same IMMATH task using the relation given below \citep{kim_trippe_2014_JKAS}
\begin{equation}\label{eq:SP-error}
\alpha_{err}(\alpha_{\nu_2,\nu_1})=\frac{1}{\log(\nu_2/\nu_1)}\times\left[\frac{\sigma_{\nu_1}^2}{I_{\nu_1}^2}+\frac{\sigma_{\nu_2}^2}{I_{\nu_2}^2}\right]^{\frac{1}{2}}
\end{equation}

with $I$ as the total intensities at the respective frequencies at each pixel.

\subsubsection{Computing magnetic field and the spectral age} \label{sec: spectral_age}

The assumption that the energy density of cosmic rays and the magnetic fields are almost the same when the magnetised plasma in a synchrotron source is near its minimum energy state, allows one to estimate the strength of magnetic fields in such systems directly from their synchrotron radio emissions \citep{1980_miley}. Such estimate of magnetic field is often termed as the equipartition magnetic field and is calculated using the relation

\begin{equation}
\label{eq: equipartition magnetic field}
    B_{eq} = 7.91  \left[ \frac{1+k}{(1+z)^{\alpha - 3}}  \frac{S_0}{\nu_0^{\alpha} \theta_x \theta_y s} \frac{\nu_u^{\alpha+\frac{1}{2}} - \nu_l^{\alpha+\frac{1}{2}}}{\alpha+\frac{1}{2}}\right]^{\frac{2}{7}} \rm{\mu G},
\end{equation} 

\noindent Here, $S_0$ is the flux density (in mJy) at the observing frequency ($\nu_0$), $k$ is the ratio of the energy content in relativistic protons to that in electrons, $\nu_l$ and $\nu_u$ are the lower and upper-frequency limits for computing the integrated radio luminosity, $s$ is the path length through the source along the line of sight (in kpc), and $\theta_x$ and $\theta_y$ are source extents in two directions, measured in arcsec. For our case, we have adopted $k=1$, $\nu_l =0.01$ GHz and $\nu_u = 100$ GHz and $\theta_x$, the size and $\theta_y$, the path length have been assumed to be the same as the average size of the source.

To produce a spectral age map for the diffuse emission observed in A980, we fit JP model using the BRATS software package \citep{harwood_2013MNRAS}. The maps used are at 144 MHz, 325 MHz and 1.5 GHz and the computation was done pixel-by-pixel in the region above 2$\sigma$ of 1.5 GHz map. As input parameter, we use the injection index ($\alpha_{inj}=0.7$) computed using the BRATS software itself and an equipartition magnetic field ($B_{eq}=2.6~\mu$G) estimated using eq.~\ref{eq: equipartition magnetic field} for the region A.

For the regions where data are available at only two frequencies, the spectral age has been calculated using the procedure described in \citep{2004AA_jamrozy}. In this model, with the assumption that a source has a uniform magnetic field, and ignoring any expansion losses, the age of a radio source with a spectral break frequency $\nu_{br}$ is given by  
\begin{equation}
    \tau = 1.59 \times 10^3  \nu_{br}^{-0.5} \frac{B_{eq}^{0.5}}{B_{eq}^2+B_{CMB}^2} \;\;\rm{Myrs},
\end{equation}
Here, $B_{eq}$ is the equipartition field, $B_{CMB} (\mu G) = 3.18 (1+z)^2$ is the magnetic field equivalent to the cosmic microwave background \citep{Longair_2011} and $\nu_{br}$ is the break frequency (in GHz). Obtaining precise break frequency from models would require observations at several well-spaced frequencies, in the absence of which we assume the break frequency as $\nu_{br} = 144$ MHz, the lowest observing frequency available in our study.

\subsection{The X-ray data}\label{sec:Xray-data}

The level-1 event file of A980 from {\it Chandra} data archive (14\,ks; ObsID 15105) was reprocessed following the standard data-reduction routine of CIAO\footnote{\color{blue}{{http://cxc.harvard.edu/ciao}}}~4.11 and employing the latest calibration files CALDB 4.8.3. Events were screened for cosmic-rays using the ASCA grades, and were reprocessed by applying the most up-to-date corrections for the time-dependent gain change, charge transfer inefficiency, and degraded ACIS detector quantum efficiency. Periods of high background flares exceeding 20\% of the mean background count rate were identified and removed using the {\tt lc\_sigma\_clip} algorithm. This yielded 13.5\, ks of net exposure time. The standard blank-sky data sets\footnote{\color{blue} {http://cxc.harvard.edu/ciao/threads/acisbackground/}} were processed and re-projected to the corresponding sky positions and normalized to match the count rate in $10-12$~keV energy range. Point sources across the ACIS field were identified using the CIAO tool {\tt wavdetect} and removed. Thereafter, exposure correction was applied on the cleaned X-ray image (i.e free from point sources and flares), using the mono-energetic exposure map created at 1~keV.

In addition, a 2D temperature map has been made using the contour-binning algorithm by \cite{sanders2006}. This algorithm generates a set of regions following the distribution of surface brightness such that each region has nearly the same signal-to-noise ratio. A total of 9 regions were thus delineated, each with a signal-to-noise ratio of 20 ($\sim400$ counts per bin), from the $0.5-3.0$~keV image. We then extracted the X-ray spectra for these regions and fitted them with the model {\tt TBABS}$\times${\tt APEC} using the method described in \ref{apdx:glob-xray}. Since the photon counts in individual patches are not statistically robust, we estimated the metallicity in circular annuli and used the values corresponding to different patches, for computing their temperatures. The estimated metallicity varies from the peak of 0.68~$\Zsun$ at the central annulus to 0.11~$\Zsun$ at the outermost annulus of the ICM.

\subsubsection{The global X-ray properties}\label{apdx:glob-xray}
The global properties of the cluster X-ray emission were determined by extracting a 0.5-8\,keV spectrum of the X-ray photons from within a 150\arcsec\,(408\,kpc) circular region centred on the X-ray peak of the cluster. We excluded the central 2\arcsec region, as well as the regions corresponding to the obvious point sources detected within the chip. A corresponding background spectrum was extracted from the normalized blank sky frame and appropriate responses were generated. The spectrum was binned such that every energy bin contains at least $\sim$20 counts and was then imported to {\tt XSPEC 12.9.1} for fitting, adopting the $\chi^2$ statistics. The combined spectrum was fitted with an absorbed single temperature {\tt APEC} model with the Galactic absorption fixed at $N_{H}^{Gal} = 9.16\times10^{19}{\rm~cm^{-2}}$ \citep{2005A&A...440..775K}, letting the temperature, metallicity and normalization parameter to vary. The best-fit resulted in the minimum $\chi^2=219.01$ for $210$ degrees of freedom. \\

\section{Results and Discussion}
\subsection{Multi-waveband view of the cluster A980}\label{sec:radio_complex}

\begin{figure*}[!ht]
    \begin{tikzpicture}
    \node(a){\includegraphics[width=0.46\textwidth]{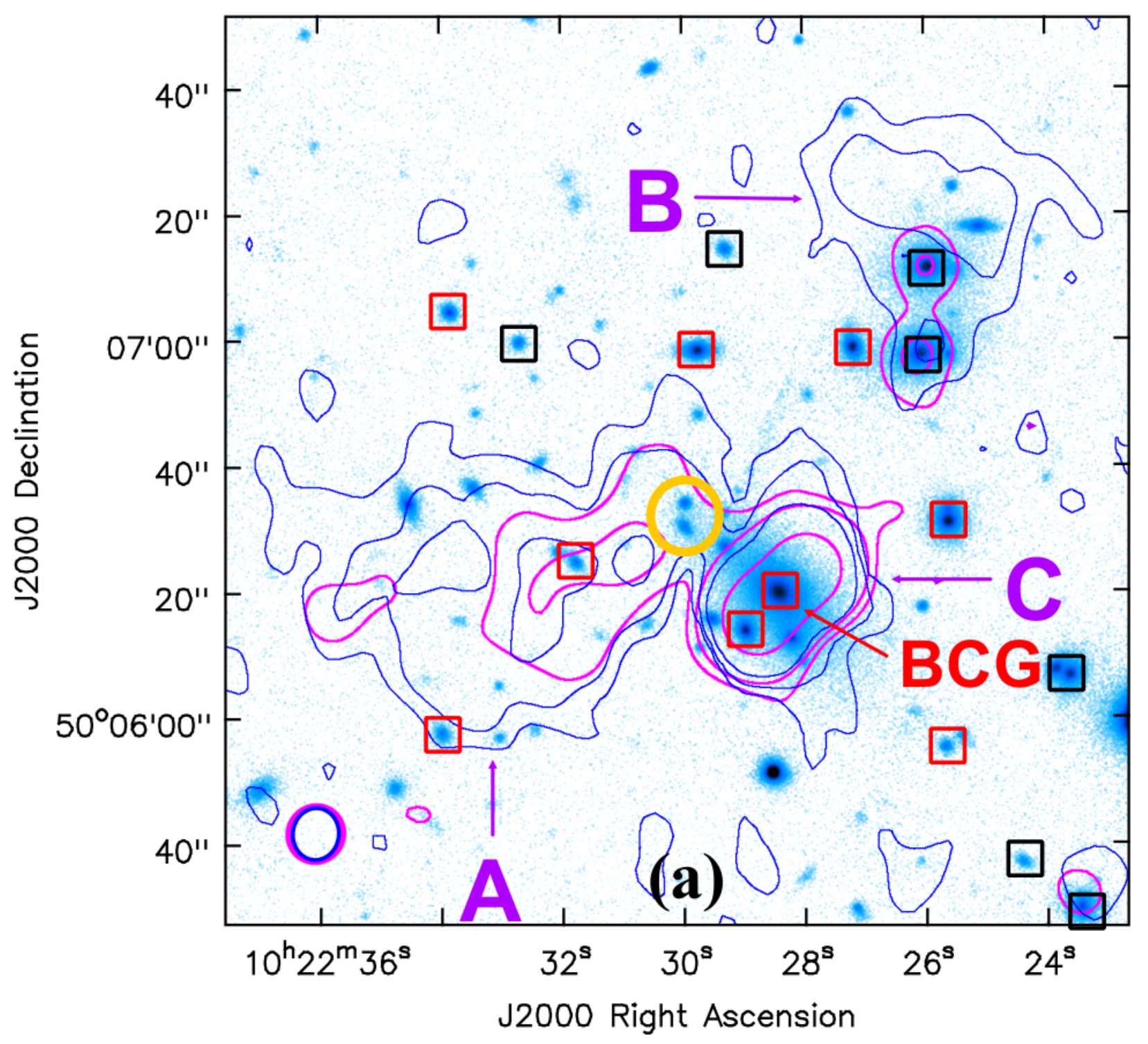}};
    \node at (a.south east)
    [
    anchor=center,
    xshift=-10mm,
    yshift=13.5mm
    ]
    {
    \includegraphics[width=0.12\textwidth]{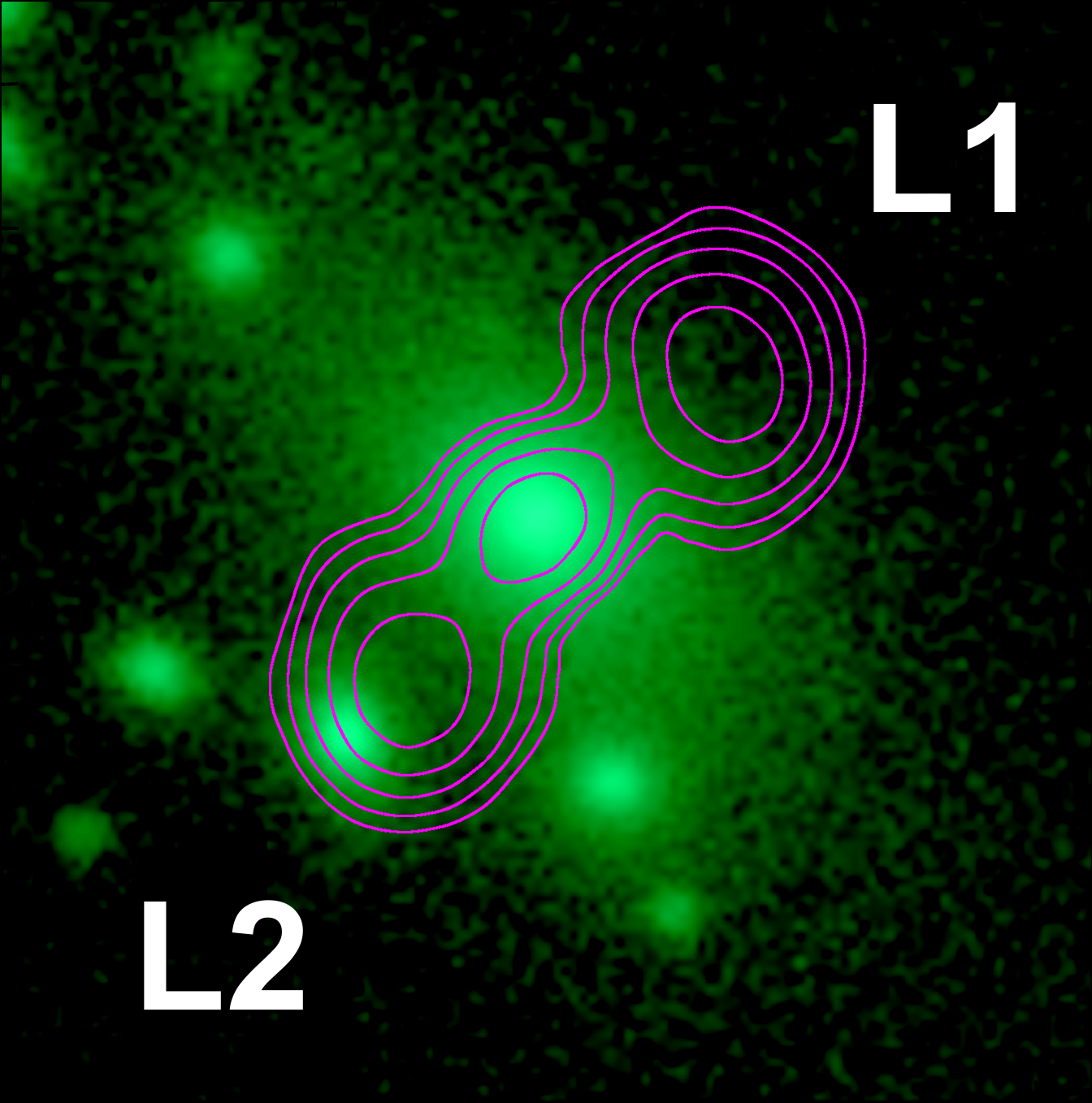}
    };
    \node at (a.north west)
    [
    anchor=center,
    xshift=29mm,
    yshift=-13.5mm
    ]
    {
    \includegraphics[width=0.12\textwidth]{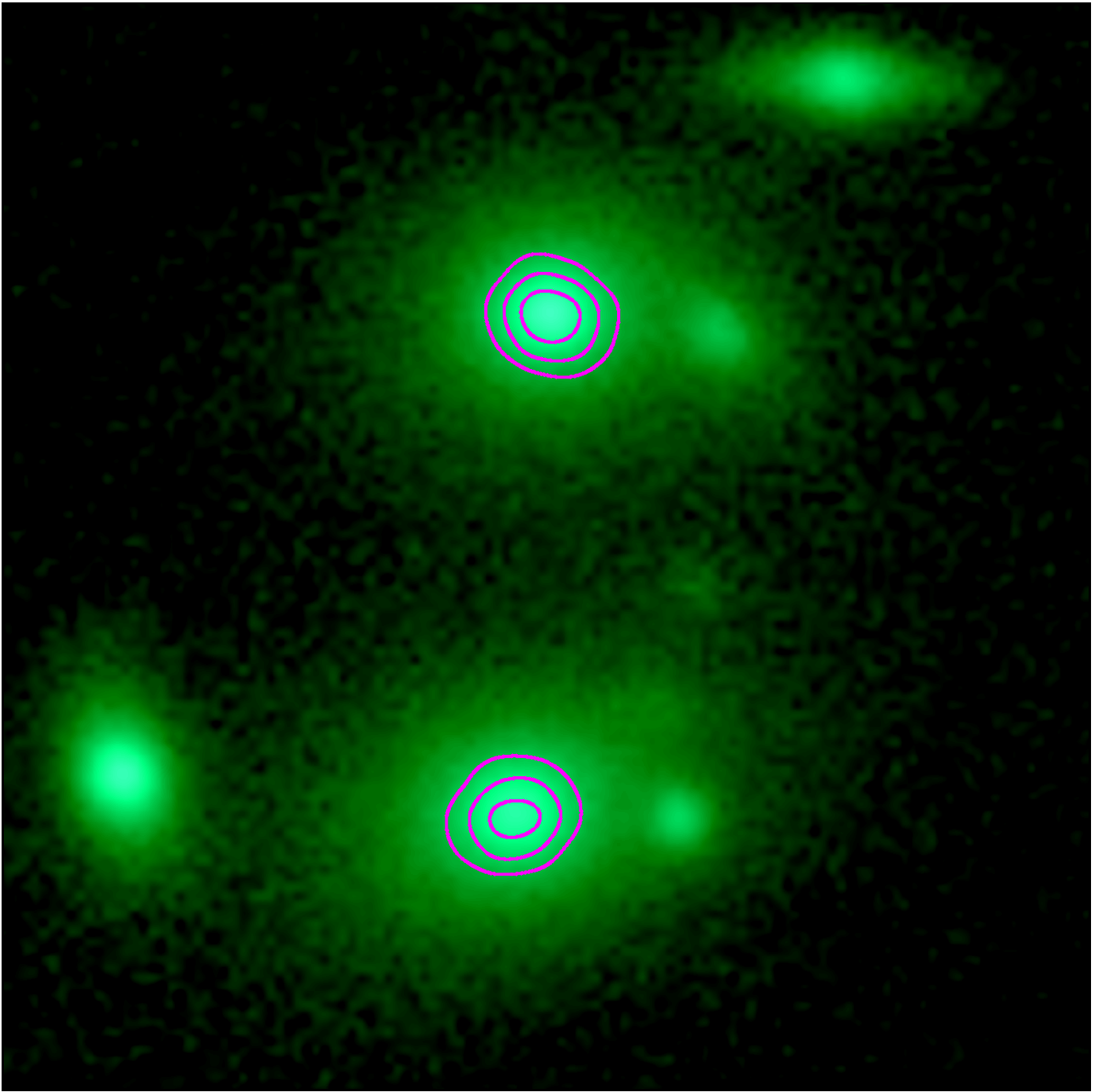}
    };
    \end{tikzpicture}
    \includegraphics[width=0.45\textwidth]{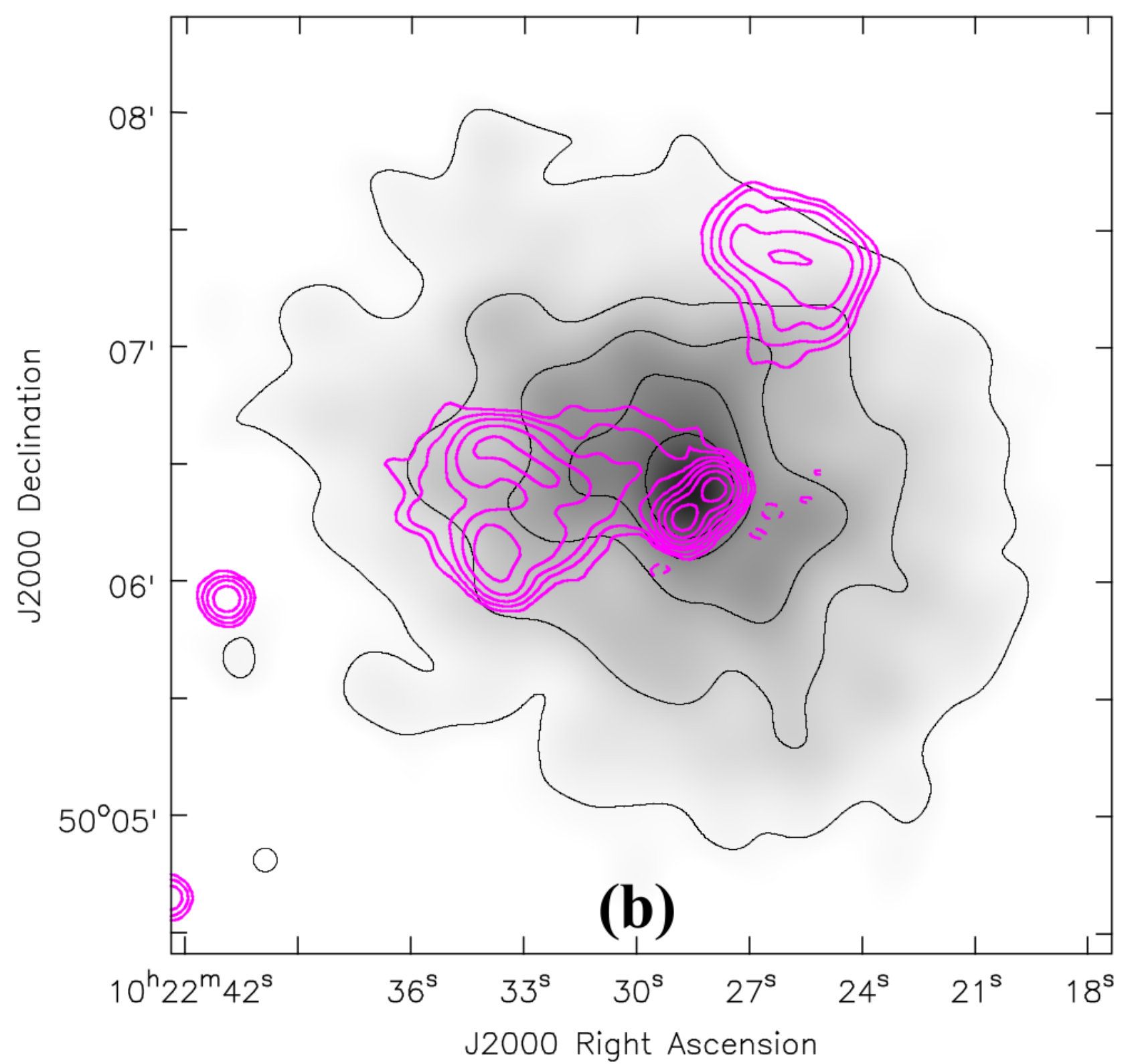}
    \caption{\textbf{Panel~(a)} Magenta contours of VLA L-band map at $ 3, 9, 81\times\sigma=25~\mu$Jy/beam (FWHM = $9.0\arcsec\times8.7\arcsec$) and blue contours (at $ 3, 9, 27 \times \sigma=100~\mu$Jy/beam) of the GMRT-325 MHz map (FWHM = $8.2{\arcsec}\times7.3{\arcsec}$) are overplotted on Pan-STARRS `i' band optical image. Among the galaxies with known spectroscopic redshifts, the more likely cluster members ($z=0.1582\pm0.0035$; red squares) and the remaining galaxies ($z<0.1547$ or $z>0.1617$; black squares) in the vicinity of the radio contours are shown. 
    The southern object within the orange circle is discussed in sect.~\ref{sec: parent_galaxy}.  The south-west and north-east insets show the high-resolution VLA contours of C and B in magenta colour, respectively at $[3, 6, 12, ....] \times \sigma$ and $[3, 5, 7] \times \sigma$ (where, $\sigma=30~\mu$ Jy/beam and FWHM=$2.8\arcsec\times2.5\arcsec$), superimposed on the same optical image. \textbf{Panel~(b)} The gray colour {\it Chandra} X-ray map (with black contours at $(2.68, 5.76, 8.80, 12.0, 15.0) \times 10^{-9}$ counts/cm$^2$/s) is over plotted with the LoTSS-2 144 MHz contours (magenta) at $-3,5,10,20,...\times 150~\mu$Jy/beam (FWHM= $6\arcsec\times6\arcsec$). \label{fig:LoTSS_GMRT_optical_xray} } 
\end{figure*}

\begin{figure*}[!ht]
    \begin{tikzpicture}
    \node(a){\includegraphics[width=0.49\textwidth]{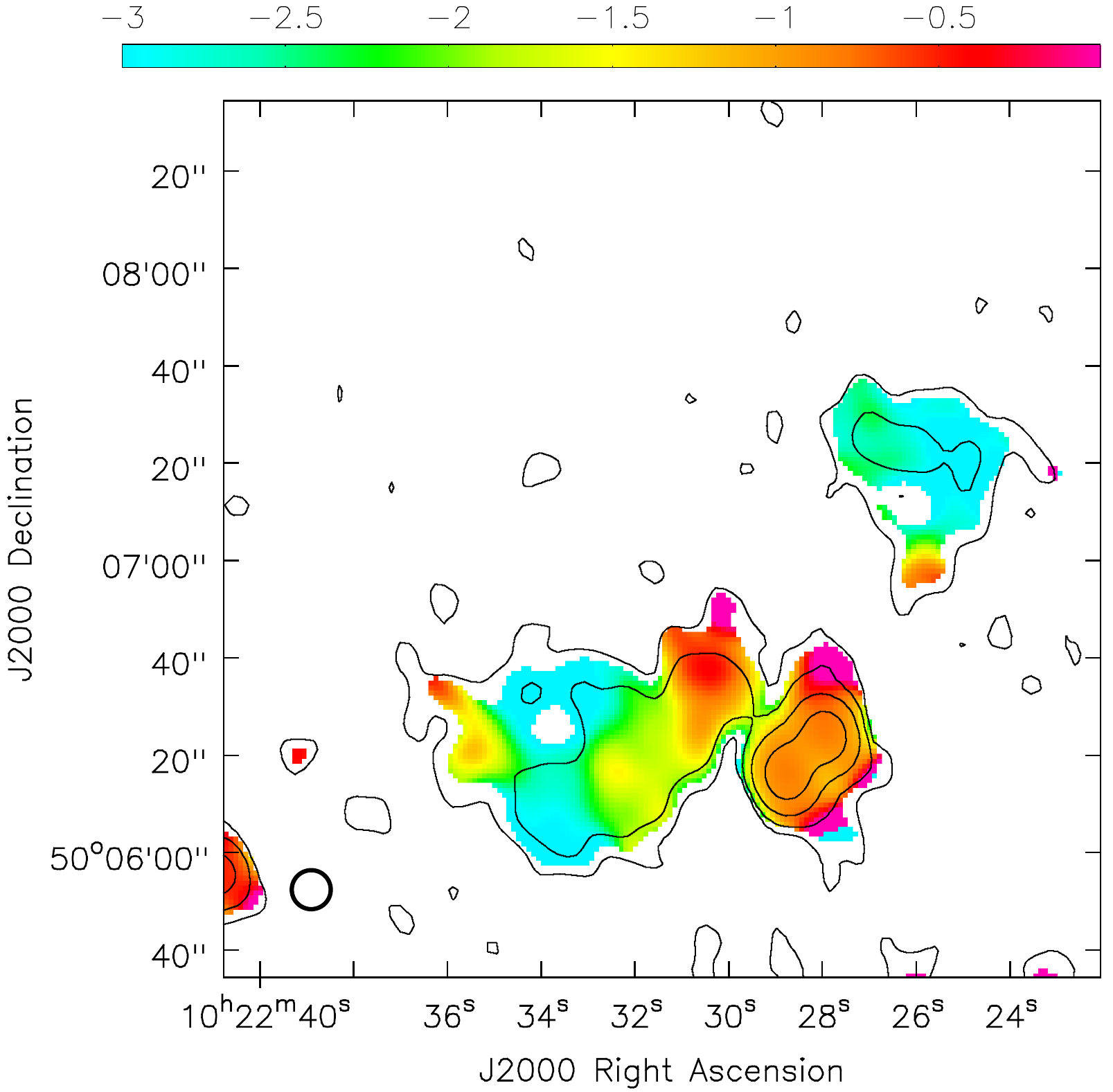}};
    \node at (a.north west)
    [
    anchor=center,
    xshift=24mm,
    yshift=-29mm
    ]
    {
    \includegraphics[width=0.21\textwidth]{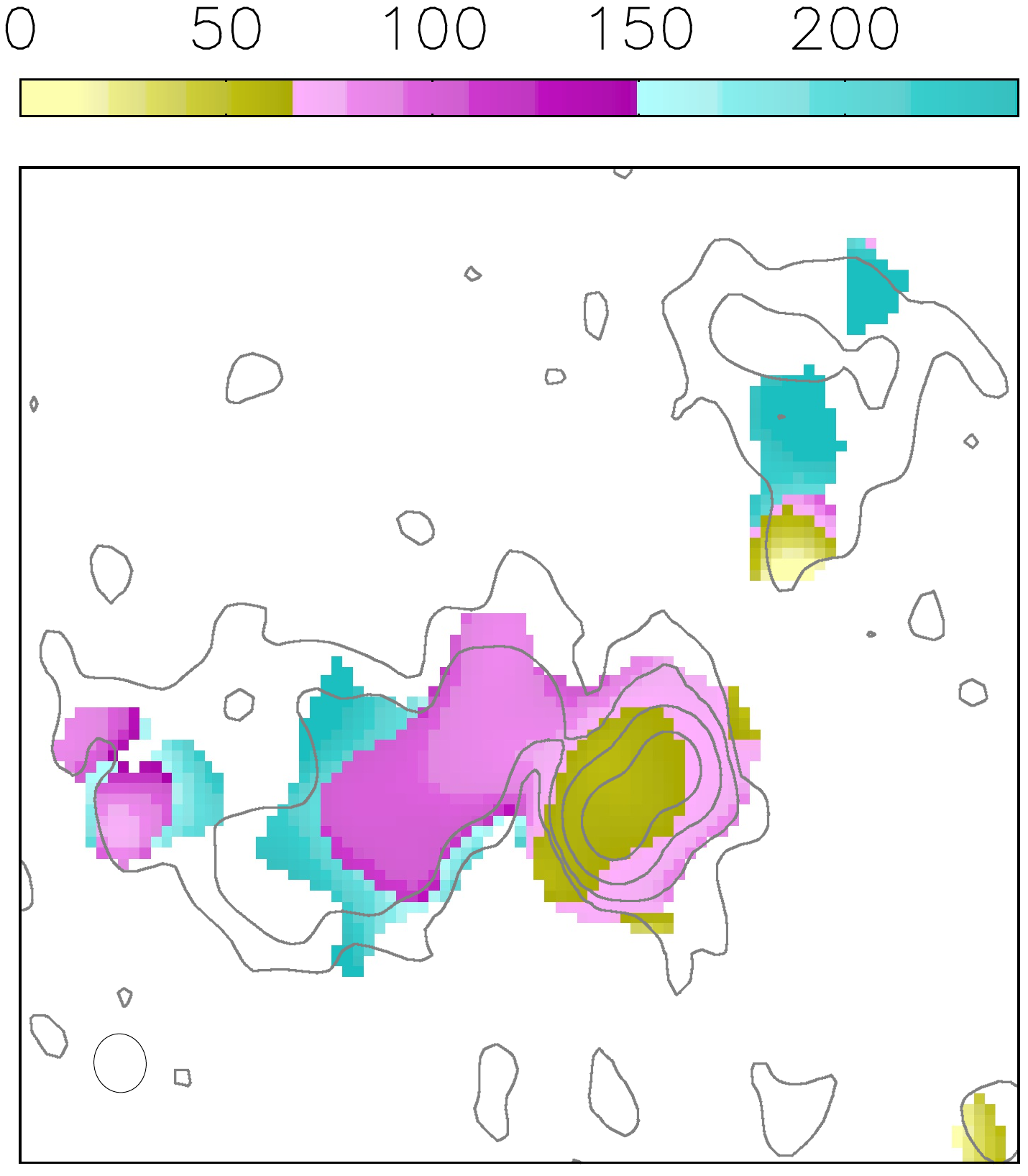}
    };
    \end{tikzpicture}
    \includegraphics[width=0.49\textwidth]{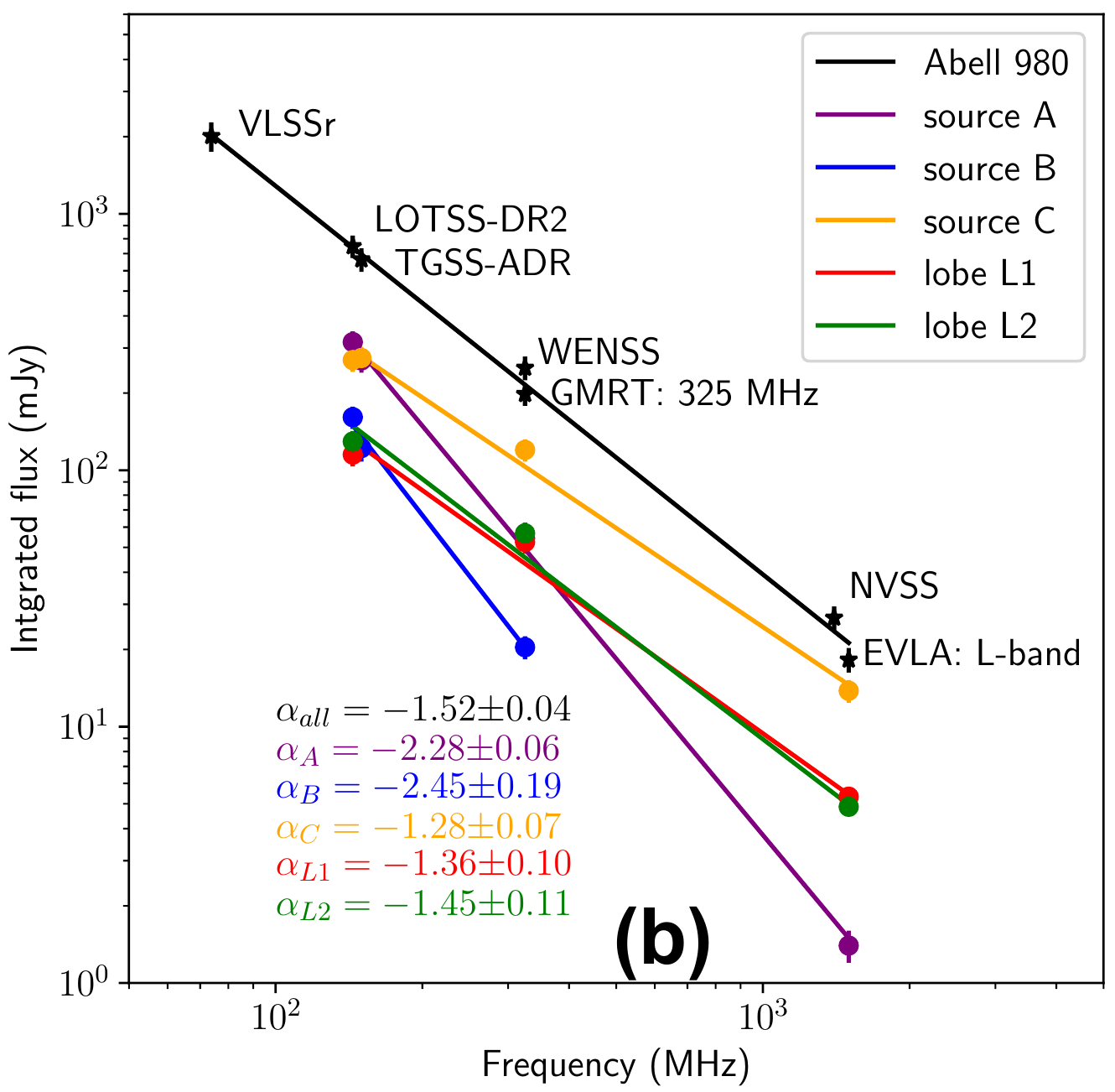}
    \caption{{\bf Panel~(a)} Spectral index distribution, derived by combining the GMRT (325~MHz) and LoTSS-2 (144~MHz) maps (see corresponding error map in \ref{apdx:radio-spectrum}). The colour map of spectral age produced by BRATS software (see sec.~\ref{sec: spectral_age}) is shown within the inset (in units of Myr) and the corresponding error map in \ref{apdx:spectral_age_error}. Contours of the low-resolution map at 325 MHz are plotted at $3, 12, 48, 192 \times 100 \mu$Jy/beam over both the maps. {\bf Panel~(b)} Spectral plots for the integrated emission, and for the different components mentioned in the legend.}\label{fig:Spectral_map_plot} 
\end{figure*}

\begin{figure*}[!ht]
\begin{tikzpicture}
    \node(b){\includegraphics[width=0.47\textwidth]{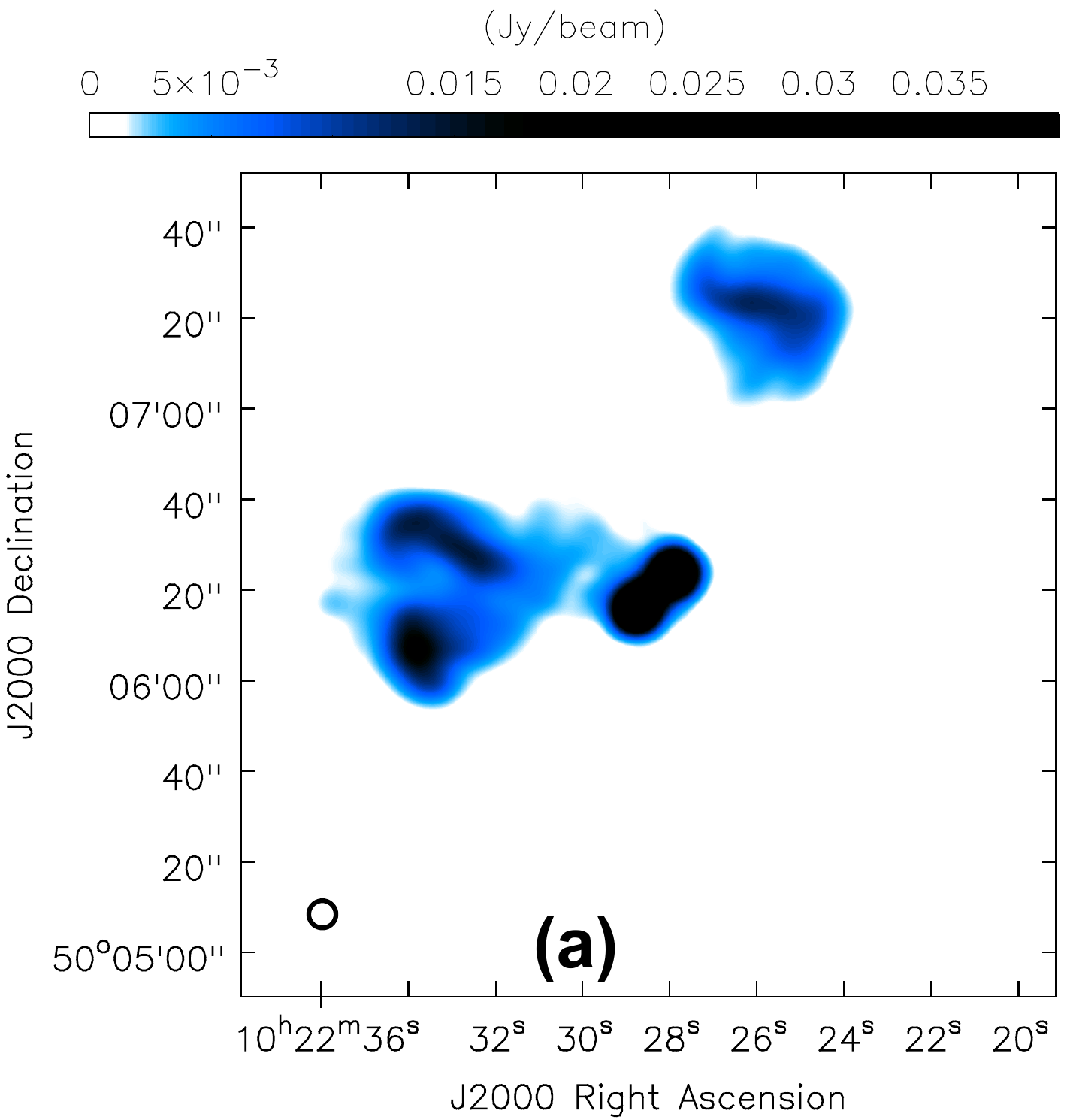}};
    \node at (b.south east)
    [
    anchor=center,
    xshift=-15.6mm,
    yshift=24mm
    ]
    {
        \includegraphics[width=0.14\textwidth]{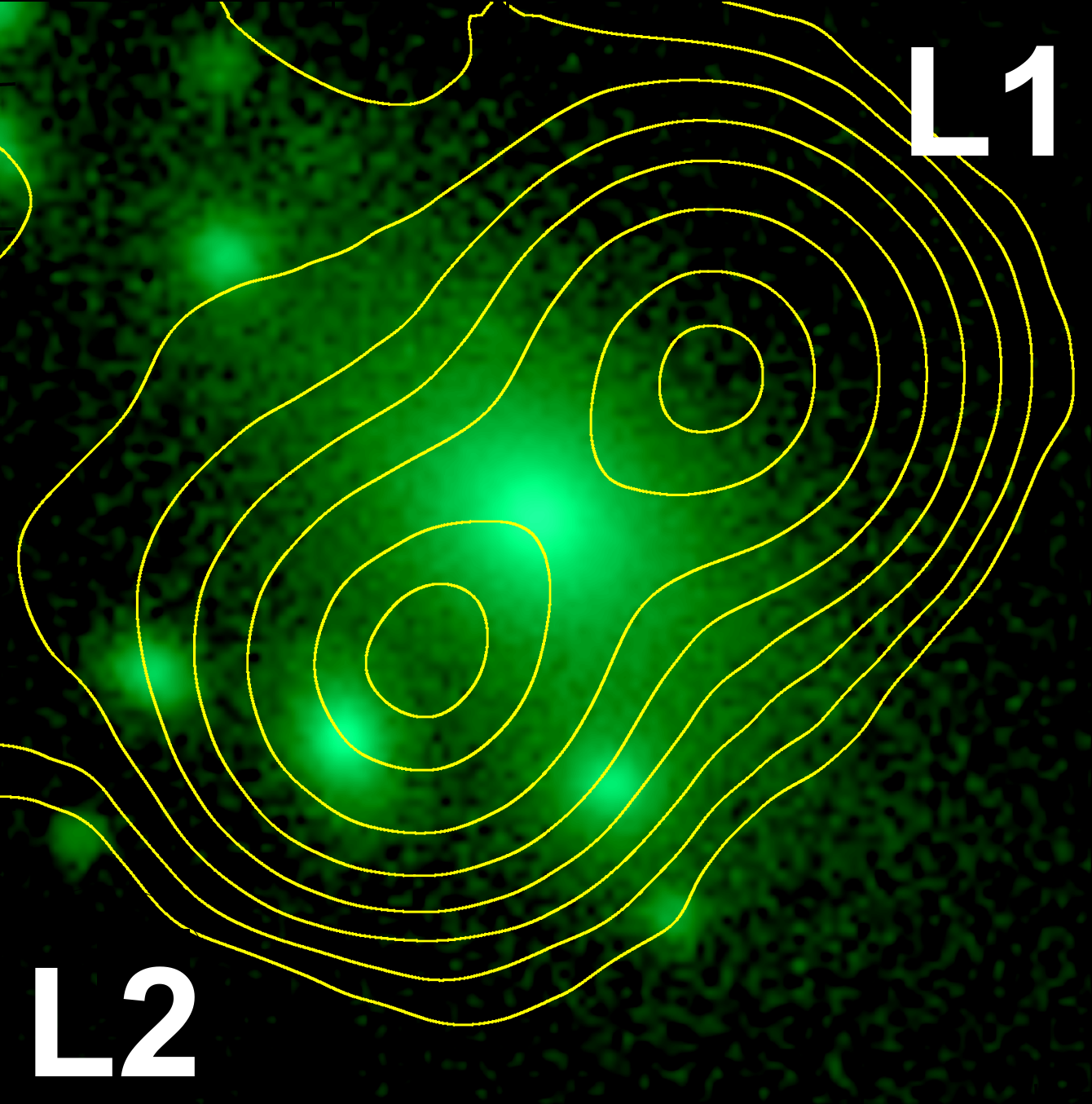}
    };
    \end{tikzpicture}
    \begin{tikzpicture}
    \node(a){\includegraphics[width=0.48\textwidth]{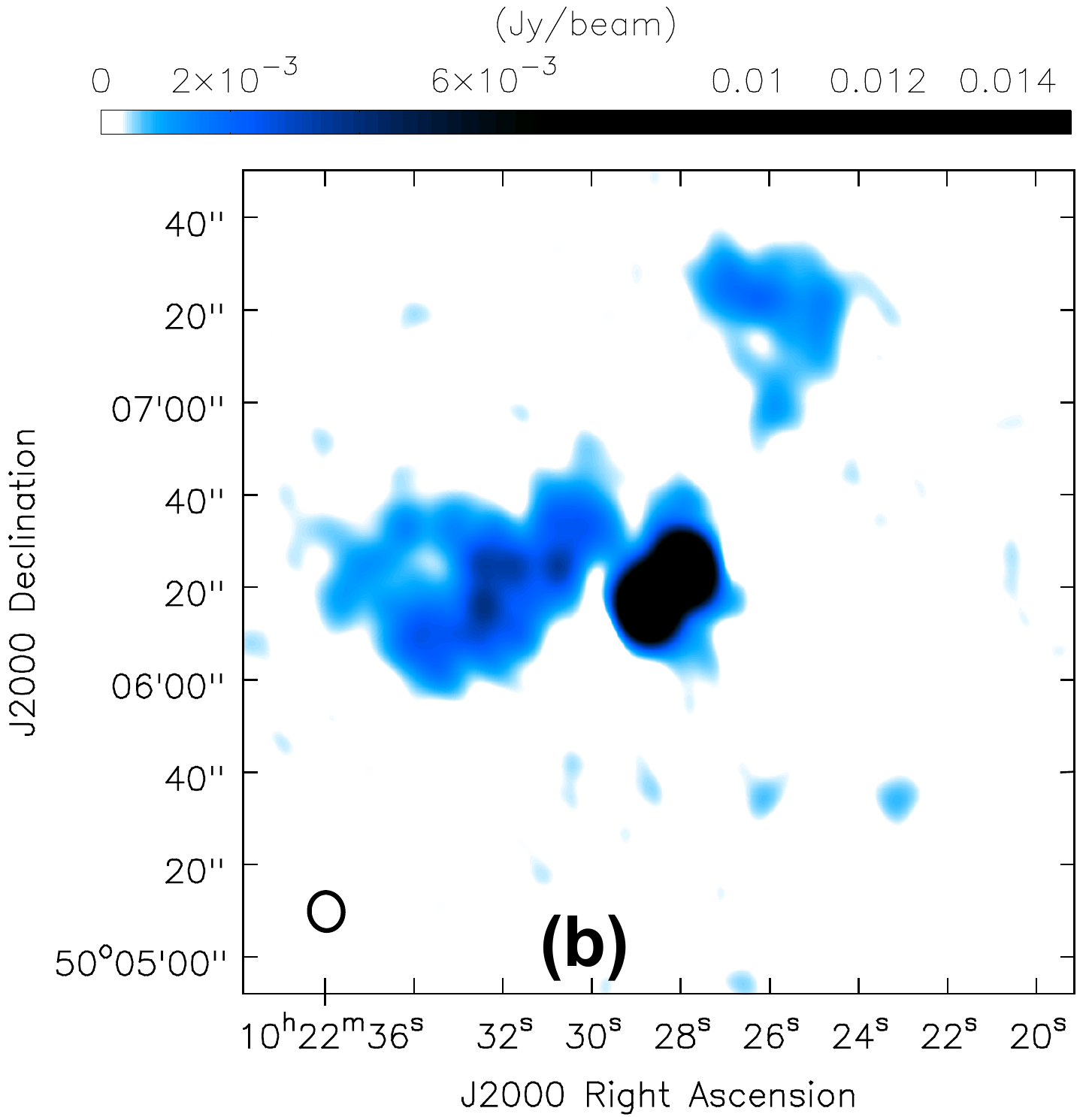}};
    \node at (a.south east)
    [
    anchor=center,
    xshift=-16mm,
    yshift=24mm
    ]
    {
        \includegraphics[width=0.14\textwidth]{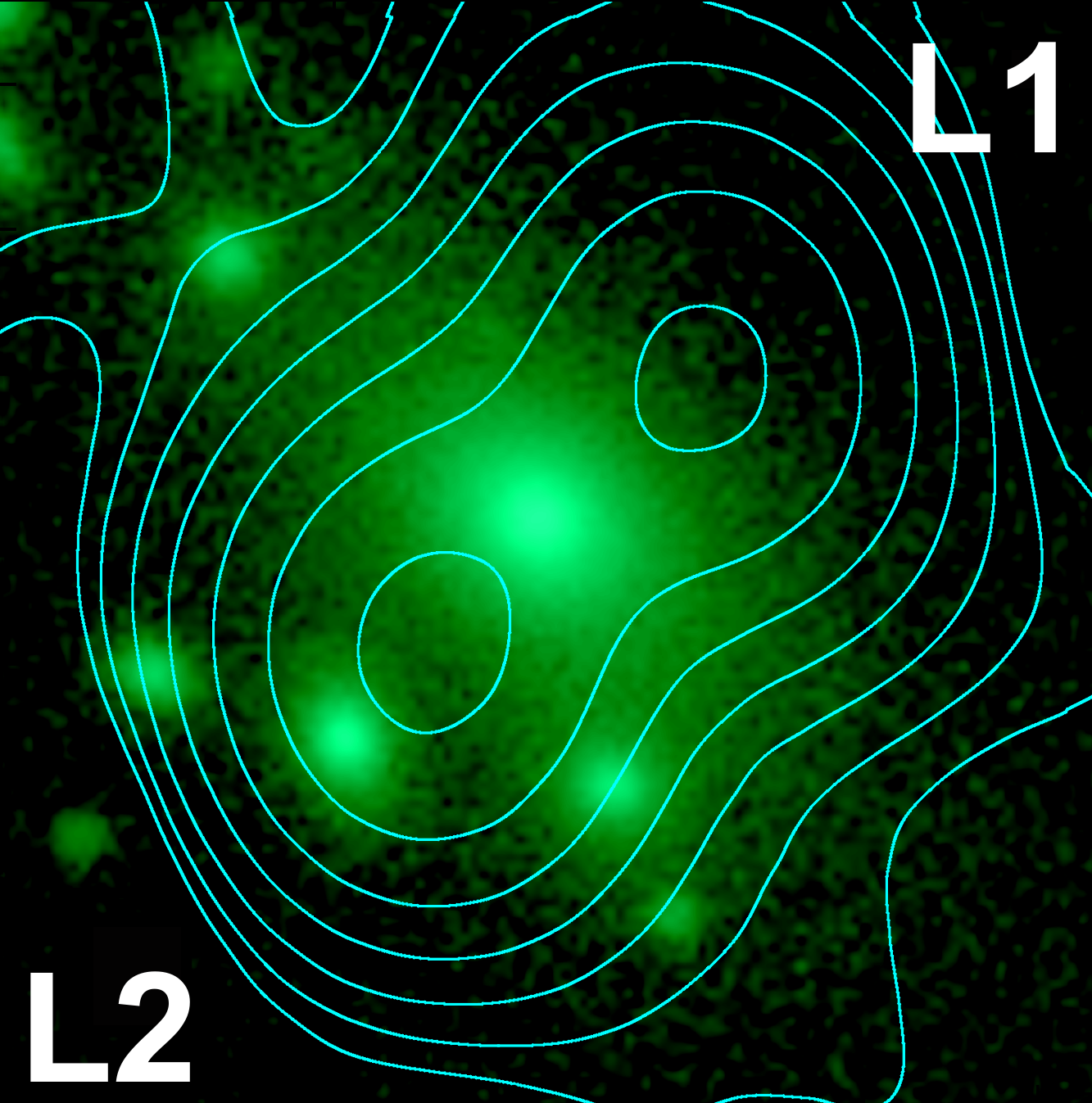}
    };
    \end{tikzpicture}

\caption{{\bf Panel~(a)}: High-resolution ({$6\arcsec\times6\arcsec$}) map from LOTSS-DR2 (144 MHz) is shown in blue colour. The inset shows the corresponding contours for the central double radio source in yellow colour, at $1, 2, 4, 8, ... \times 750~\mu$Jy/beam, over-plotted on the Pan-STARRS `i' band optical image. {\bf Panel~(b)} GMRT map (325 MHz) is shown in blue colour and the inset shows the central double source in cyan colour contours, at $1, 2, 4, 8, ... \times 650~\mu$Jy/beam} \label{fig: A980_grey_lotts2_gmrt}
\vspace{-0.1cm}
\end{figure*}

\begin{table*}
\caption{The measured parameters of the radio components and of the integrated radio emission (see Fig.~\ref{fig:Spectral_map_plot}b)} %
\centering    
\small
\begin{tabular}{lcccccccccc}
\hline
\vspace{10pt}
Source &  \multicolumn{5}{c}{Flux density (mJy)} & \multicolumn{5}{c}{Spectral index}\\
\hline
& LOTSS-2 & TGSS & GMRT & WENSS & VLA & & & & &\\
& 144 MHz & 150 MHz & 325 MHz & 325 MHz & 1.5 GHz & $\alpha_{\rm{144}}^{\rm{325}}$ & $\alpha_{\rm{150}}^{\rm{325}}$ & $\alpha_{\rm{325}}^{\rm{1500}}$ & $\alpha_{\rm{144}}^{\rm{1500}}$ & $\alpha_{int}$ \\
\hline  
A & $316\pm32$ & $268\pm28$ & $54.3\pm5.5$ & - & $1.4 \pm 0.2$ & $-2.3\pm0.2$ & $-2.1 \pm 0.3$ &$-2.4 \pm 0.2$ & $-2.3 \pm 0.1$ & - \\
B & $161\pm16$ & $122\pm14$ & $20.4\pm2.1$ & - & $<0.89$ & $-2.5\pm0.3$ & $-2.3 \pm 0.3$  & $< -2.1$  & $< -2.2$ & - \\
C & $269\pm27$ & $274\pm28$ & $120\pm12$ & - & $13.8\pm1.4$ & $-1.0\pm0.3$ & $-1.1\pm0.3$ & $-1.4\pm0.1$ & $-1.3 \pm 0.1$ & - \\
L1 & $115\pm12$ & - & $52.5\pm5.3$ & - & $5.3 \pm 0.3$ & $-1.0\pm0.3$ & - & $-1.5\pm0.1$ & $-1.3\pm0.1$ & - \\
L2 & $129\pm13$ & - & $56.8\pm5.7$ & - & $4.9 \pm 0.2$ & $-1.0\pm0.3$ & - & $-1.6 \pm 0.1$  & $-1.4 \pm 0.1$ & - \\
Integrated & $747\pm75$ & $660\pm67$ & $198\pm20$ & $251 \pm 27$ & $18.3\pm2.0$ & - & - & - & - & $-1.5 \pm 0.1$\\
\hline
\hline
\end{tabular}
\label{tab:diffuse_emissions} 
\end{table*}

\begin{figure}
\includegraphics[width=0.98\textwidth]{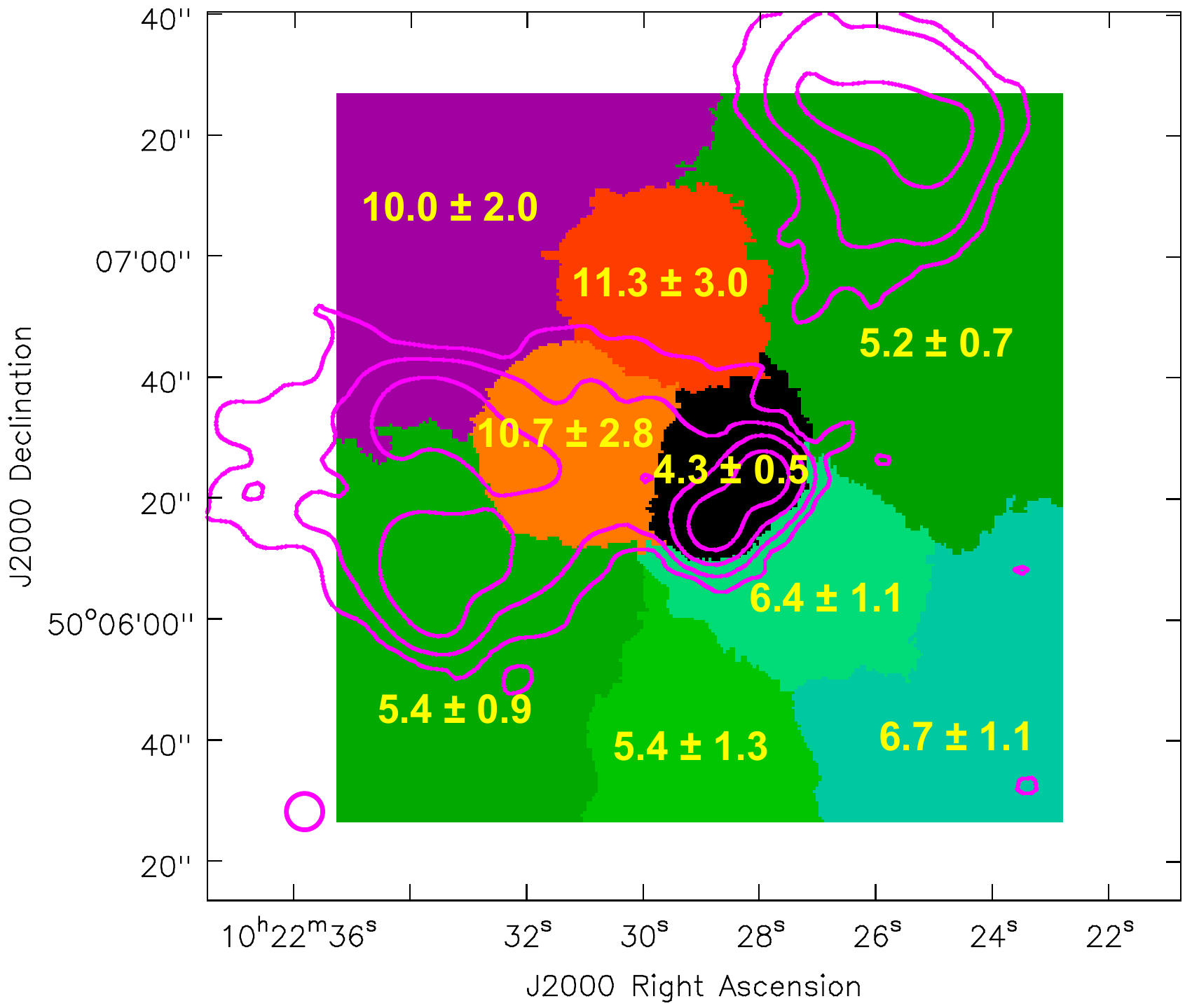}
\caption{{\it Chandra} X-ray temperature (keV) map overlaid with the radio contours (at $5,20,80,320 \times 150~\mu$Jy/beam) of the LOTSS-2 map.}\label{fig:X-ray_temp_GMRT}
\end{figure}

For a perspective of this system, we display in Figs. \ref{fig:LoTSS_GMRT_optical_xray}a~\&~\ref{fig:LoTSS_GMRT_optical_xray}b the multi-band overlays, followed by the derived radio spectral information in Figs.\ref{fig:Spectral_map_plot}a~\&~\ref{fig:Spectral_map_plot}b. 
Fig.~\ref{fig:LoTSS_GMRT_optical_xray}a shows the GMRT map (FWHM = $8.2{\arcsec} \times 7.3{\arcsec}$ at 325 MHz) in which the components A, B and C are overlaid on an optical image of the field. The adjoining overlay displays the recently published LoTSS/DR2 map (FWHM = $6\arcsec\times6\arcsec$ at 144 MHz) superposed on the {\it Chandra} X-ray image 
(Fig.~\ref{fig:LoTSS_GMRT_optical_xray}b). In Fig. \ref{fig:Spectral_map_plot}a, we present the spectral index map derived by combining the LoTSS/DR2 and GMRT maps, after matching them in angular resolution. The brighter components A and C, which are barely separated in the TGSS-ADR1 map with a 25 arcsecond beam at 150 MHz \citep{2017Intema_SPAM}, can now be clearly demarcated in the LoTSS 144 MHz map, enabling a reliable extension of their spectra down to this low-frequency (Fig.~\ref{fig:Spectral_map_plot}b; Table \ref{tab:diffuse_emissions}). Fig.~\ref{fig: A980_grey_lotts2_gmrt} shows these two radio maps in blue-colour. Lastly, Fig.~\ref{fig:X-ray_temp_GMRT} displays the LoTSS-2 high-resolution contour map overlaid on the ICM temperature map in which a hot patch (T $\sim 11$~keV) is present just to the north-east of the cool core of this cluster.

\subsection{Parent galaxy of the USS radio components A and B} \label{sec: parent_galaxy}

The radio ridge of component A, extending towards B (Figs.\ref{fig:LoTSS_GMRT_optical_xray}a \&~\ref{fig: A980_grey_lotts2_gmrt}b) strongly suggests that these two USS components are physically linked relic lobes ($\alpha<-2.0$; Fig. \ref{fig:Spectral_map_plot}b). The lack of a terminal hot spot means that the radio ridge seen in A is unlikely to be a manifestation of back-flowing synchrotron plasma. This is clearly borne out by the radio spectral gradient derived here, which shows the spectrum to become flatter going inward (Fig. \ref{fig:Spectral_map_plot}a). This provides a vital observational evidence for the hypothesis that the two lobes are aged synchrotron plasmons produced by one galaxy during its active phase and the oldest synchrotron plasma, propelled by buoyancy in the cluster potential, has now reached the ICM outskits (S22). The spectral gradient derived here shows the rising USS plasmons being trailed by younger plasma. Observational clue in support of this buoyancy interpretation \citep{Gull_1973Natur, Sakellio_1996MNRAS} comes from the good alignment between the front edge of each component with the X-rays contour at that location (Fig.~\ref{fig:LoTSS_GMRT_optical_xray}b). The observed lateral spreading of the lobes is expected to set in when a plasmon has buoyantly risen to the iso-entropy surface within the ICM (e.g., \citealt{Kaiser_2003MNRAS} ). Secondly, the spectral ageing analysis (\ref{sec: spectral_age}) gives an age of $\sim260$ Myrs for the outer parts of the relic lobes A and B, which can easily allow the proposed long buoyant drift through the ICM with an expected average speed roughly a third of the ICM sound speed \citep{2001ApJ_churazov}, c$_s$ (ICM), which is $\sim 1350$~km/s for T$_{ICM} \sim 7$~keV (\S \ref{intro}). The alternative scenario that A and B are aged radio trails of a Wide-Angle-Tail (WAT) source is less appealing because the radio contours of both A and B in the more sensitive LoTSS map (Fig. \ref{fig:LoTSS_GMRT_optical_xray}b) show no sign of trailing off towards the expected north-eastern direction. Here it may also be mentioned that the southward radio spur extending from component B is a feature unrelated to B since it is contributed by two point-like radio sources detected in the 1.5~GHz EVLA map, which coincide with the two galaxies shown in black squares in Fig. \ref{fig:LoTSS_GMRT_optical_xray}a (see, also the north-eastern inset).

In contrast to FR II sources, the spectrum is seen to become flatter going inwards along the radio lobe A (Fig.~\ref{fig:Spectral_map_plot}a). However, such a spectral gradient is seen in the cluster radio galaxy 3C 388 and explained in terms of intermittent jet activity in its host galaxy \citep{Roettiger_1994ApJ,Brienza_2020A&A}. Conceivably, the inner part of the radio ridge in component A is a manifestation of younger synchrotron plasma injected by residual low-level jet activity of the host galaxy (see, e.g., \citealt{Sabater_2019A&A}), while the wider outer part of A (and B) are the remnants from the earlier, main active phase of the host galaxy. The spectral ageing analysis using BRATS software gives $\sim 90$~Myr for the age of the synchrotron plasma near the centre, gradually rising to $>240$~Myr for the outer parts of lobe A (see inset of Fig.~\ref{fig:Spectral_map_plot}a). The regions of lobe B and the part of A, for which radio emission at only two frequencies are detected, the ages have been computed following the procedure described in \citep{2004AA_jamrozy} and adopting the JP model (\ref{sec: spectral_age}), and are found to be $255\pm45$~Myrs and $261\pm32$~Myrs, respectively. Interestingly, Fig.~\ref{fig:LoTSS_GMRT_optical_xray}a does not show even a moderately bright optical galaxy near the central location between A and B. Of the two optical objects seen within the orange circle, the northern one is classified as a star and the lower one as a galaxy (SDSS/DR7, \citealt{Abazajian_2009ApJS}), lacking radio detection (Fig.~\ref{fig:LoTSS_GMRT_optical_xray}a). The values of apparent magnitude (m$_r$) of this galaxy are given as  19.96 $\pm$ 0.05 (PANSTRASS-1, \citealt{chambers_2016arXiv} ) and 20.64 $\pm$ 0.09 (SDSS/DR7, \citealt{Abazajian_2009ApJS}). Its redshift is not known and it would be desirable to measure it, but assuming that it does lie within the cluster A980, the absolute magnitudes would be M$_r$ = - 19.44 $\pm$ 0.07 and - 18.76 $\pm$ 0.10, respectively. These can be compared with M$_r$ = - 22.8 $\pm$ 0.5 for the hosts of radio galaxies at $z <~ 0.5$, for the same cosmology as the one adopted in the present work \citep{Treves_2005ApJ}. Thus, a comparison with the above PANSTRASS-1 and SDSS/DR7 based estimates of M$_r$ implies that the galaxy within the orange circle, if indeed a member of A980, is under-luminous by 3.36-mag  and  4.04-mag, respectively, i.e., 6.6$\sigma$ and 7.9$\sigma$, where $\sigma$ is the combined rms error. This makes it a highly improbable host of the relic lobe pair A - B. Discarding this unlikely host, the plausibility favours the scenario 
(sec.~\ref{intro}) that the relic lobes A and B were both created by the BCG during its previous major episode of jet activity when it was near the central location between A and B. Since then, the galaxy has drifted towards the cluster centre (BCG in Fig.\ref{fig:LoTSS_GMRT_optical_xray}a) and entered a new phase of intense jet activity which is identified with the creation of the smaller double radio source with an active central core (L1, L2 in the inset of Fig. \ref{fig:LoTSS_GMRT_optical_xray}a). Thus, in this scenario, the buoyant rise of the diffuse relic lobes A and B within the ICM, following the near-cessation of intense jet activity in their parent galaxy (BCG), and the migration of this galaxy towards the cluster’s gravitational centre are two concurrent processes. Besides the obvious fact that the BCG is massive enough to engender a powerful large double radio source, this scenario is supported by the following additional arguments:

(a)  The radial migration of the active galaxy (BCG) to the cluster centre, as required in the present scenario, is the natural trajectory for the galaxy to take (a requirement of migration in the opposite direction would have rendered the scenario untenable). Note also that only a modest average speed of $\sim 725$ km/s is needed for the galaxy to traverse the (projected) radial distance of $\sim 70$~kpc from the A — B axis to its present location near the cluster centre, during the estimated spectral age ($\sim95$~Myr) of the plasma around the mid-point between the relic lobes A and B (sect. \ref{sec:radio_complex}; Fig.~\ref{fig:Spectral_map_plot}a). Such a speed would be entirely consistent with the observed velocity dispersion of $1033$~km/s for the galaxies in this cluster \citep{Rines_2013ApJ}.

(b) An important piece of supporting evidence contributed by the LoTSS map (Fig. \ref{fig: A980_grey_lotts2_gmrt}a) is the faint radio spur emerging orthogonally from the A -- B axis and connecting to the BCG, thus tracing the proposed radial trajectory for the drift of the BCG into the cluster centre. Although a hint of this spur is visible in the GMRT map  (Fig. \ref{fig: A980_grey_lotts2_gmrt}b), its independent detection in the LoTSS map is crucial, for this confirms the reality of this weak feature and thus provides an important clue. Quite plausibly, this radio spur arises from some of the magnetised relativistic plasma of the lobe A being dragged along in the wake of the moving BCG, with additional contribution coming from the residual low-level activity in the moving BCG (see,e.g., \citealt{Sabater_2019A&A}).

(c) The A -- B axis is parallel to the axis of the young double (L1, L2). This accords with the frequently observed near-constancy of the jet direction between successive episodes of jet activity in DDRGs (e.g., \citealt{Saikia_2009BASI}). A similar case for near-constancy of the jet direction has been made for X-shaped radio galaxies, as well (\citealt{Cotton_2020MNRAS} and references therein).

(d) The hypothesis that the USS sources A and B have a common host galaxy, is also supported by the presence of the extra-hot patches of the ICM, seen between A and B (shown in red/orange colour in Fig. \ref{fig:X-ray_temp_GMRT}). The ICM temperature at these patches is $\sim 10\%$ higher than the surroundings. Such a temperature enhancement associated with powerful jet activity in clusters has been observed in numerical simulations \citep{Raouf_2017MNRAS}. Since both hot patches fall outside the cool core, their high temperature is expected to have persisted over the age of the relic lobes  A and B.

The recurrence of jet activity in the BCG, which is germane to the present scenario, is also entirely consistent with the notion that cool cores of clusters are propitious sites for triggering jet activity (e.g., \citealt{Burns_1990AJ,Fabian_2006MNRAS,Mittal_2009A&A}).  A plausible reason is that warm extended filaments and cold clouds condensing out of the ICM would accrete steadily onto the SMBH in the cluster central galaxy  
(e.g., \cite{Pizzolato_2005ApJ,Sharma_2012MNRAS,Voit_2015ApJ}). Moreover, their accretion rate can be boosted by almost two orders of magnitude over the Bondi prediction, as inelastic collisions between the gas clouds in the central region cause angular momentum cancellation, triggering a fresh episode of jet activity in the dominant central galaxy \citep{Gaspari_2013MNRAS}.\par 

Purely from an observational standpoint, the most dramatic manifestations of recurring jet activity are the so-called, “double-double” radio galaxies (DDRGs) in which the host galaxy is straddled by an inner (younger) and an outer (older) pair of radio lobes (e.g., \citealt{Schoenmakers_2000MNRAS,Saikia_2009BASI,Morganti_2017NatAs}). Still, rarer manifestations are the so-called “triple-double radio sources” \citep{Brocksopp_2007MNRAS} and the unique case where such a morphology is associated with the galaxy J140948.85-030232.5, a spiral within a galaxy cluster (`SPECA', \citealt{Hota_2011MNRAS}). The novel aspect of the source in A980 is the large positional shift of the parent galaxy between its major episodes of jet activity. Consequently, the two laterally well-separated pairs of old (A, B) and young (L1, L2) radio lobes, whose axes are parallel but which would customarily be deemed as two unrelated double radio sources, are actually the ‘outer’ and ‘inner’ lobe pairs of a single radio galaxy, which have lost their usual collinearity and got detached from each other owing to a lateral motion of the parent galaxy. Very likely, in this case of a {\it detached} DDRG (dDDRG), both the large lateral displacement of the parent galaxy between its activity episodes and the prolonged preservation of the identity and detectability of the older lobe pair, have been facilitated by their location inside a relaxed cluster.

Although we have argued that the radio components A and B are physically associated as a pair of fossil radio lobes of a single parent galaxy, a potential alternative scenario would be that A and B are independent radio sources. Under this hypothesis, the parent galaxy of B might be one of the two ellipticals coinciding with its south-ward radio spur, each of which has a distinct radio counterpart detected in the high resolution EVLA map, as seen within the north-eastern inset to Fig.~\ref{fig:LoTSS_GMRT_optical_xray}a. At the same time, the component A may be associated with the triple radio source C, together forming a Narrow-Angle-Tail (NAT) radio source hosted by the BCG.  Although, this hypothesis cannot be ruled out, it seems improbable in view of the following radio indicators:

(i) The radio component A shows a prominent elongation towards B and, moreover,  the two have similarly ultra-steep radio spectra (Fig.~\ref{fig:Spectral_map_plot}a) and diffuse radio morphology (Figs.~\ref{fig:LoTSS_GMRT_optical_xray}a and \ref{fig: A980_grey_lotts2_gmrt}b). All this suggests a common origin of A and B.

(ii) Far from being continuous, the radio 'connection' between A and C is very faint, almost like a discontinuity in surface brightness (Fig.~\ref{fig: A980_grey_lotts2_gmrt}b).

(iii) Instead of an expected steady spectral steepening from C towards A, a significant spectral index discontinuity is observed between them. The faint radio spur linking A and C has $\alpha=-1.85\pm0.18$ and is flanked by regions of flatter spectra ( $\alpha=-1.00 \pm 0.14$ to the north and $-0.97\pm0.05$ to the south, see Fig.~\ref{fig:Spectral_map_plot}a). It is conceivable that the flatter spectrum region to the north is due to a better confinement of the radio plasma due to the very high temperature of the ICM near that location (Fig.~\ref{fig:X-ray_temp_GMRT}). However, an expected surface brightness enhancement due to the better confinement is not observed.

(iv) In the moving BCG scenario (also essential for the NAT interpretation), radial infall is the natural trajectory for the BCG to take for arriving at its present location near the cluster centre. Any radio trail of the BCG would hence be expected to be elongated in the radial direction, whereas the component A extends almost orthogonally to that direction (Fig.~\ref{fig: A980_grey_lotts2_gmrt}b).

\subsection{The other known dDDRG candidates} \label{sec: DDRG}

{\bf 3C 338:} This $\sim 80$~kpc long radio source, hosted by the multi-peaked cD galaxy in Abell 2199, consists of (i) a small double radio source having a twin-jet emanating from a radio core, and (ii) an offset pair of much larger diffuse USS lobes joined by a collimated jet-like radio feature with $\alpha \simeq -1.9$  \citep{Burns_1983ApJ, Ge_1993AJ}. These authors have considered the possibility of this USS lobe pair being a relic from past activity in the galaxy which has since moved laterally northward and produced the younger lobe pair. However, they have also expressed caution about this scenario on account of the high brightness and strong collimation of the USS radio feature, the putative `relic jet'. In fact, this feature bears a striking resemblance to the recently discovered `collimated synchrotron threads’ in the WAT radio source ESO 137-006 \citep{ramatsoku_2020A&A}.  

{\bf 4C 35.06 (B2 0258+35B):} The complex radio morphology of this source in the cluster Abell 407 ($z$ = 0.047) has been unravelled through the radio imaging with VLA (1.4/4.9 GHz)/LOFAR (62 MHz) \citep{2015A&A_shulevski} and GMRT (150/235/610 MHz) \citep{2014ASInC_Biju, 2017MNRAS_biju}. The cluster core hosts a uniquely tight grouping of nine ellipticals (`Zwicky’s Nonet’) confined within just $\sim 50$ kpc. Two of those ellipticals are identifiable with two double radio sources of very unequal extents, whose axes are parallel but offset by $\sim 11$ kpc \citep{2017MNRAS_biju}. The larger radio source comprises a 400 kpc long pair of helical jets/lobes whose spectral index ($\alpha$) gets as steep as -2 in the outer parts. Although its proposed host galaxy is 3- 4 magnitudes fainter than typical hosts of large radio galaxies, this mass deficit has been attributed to a severe tidal stripping of its stellar envelope due to the uniquely dense galaxy environment of the Nonet in which this galaxy is located (\citealt{2014ASInC_Biju, 2017MNRAS_biju}; see, also \citealt{2015ApJ_Graham_scott}). 
In contrast, \citet{2015A&A_shulevski} have asserted that both double radio sources have been created by the same luminous elliptical UGC 2489 which now exhibits a VLBI radio core and is seen to coincide with the smaller double source. In their scheme, the larger radio source got created $\sim35$~Myr ago in a previous active phase of this massive elliptical, which has since moved laterally by $\sim 11$ kpc to its present location and entered a new active phase, resulting in the smaller double radio source. However, this interpretation remains uncertain due to the afore-mentioned debate over the host galaxy of the larger double radio source. Furthermore, the nature of the smaller `double radio source' is itself somewhat unclear, since each of its radio `lobes' coincides with an elliptical galaxy belonging to the Nonet (Fig. 1 of \citealt{2015A&A_shulevski}; Fig. 3 of \citealt{2017MNRAS_biju}). 

\vspace{-0.4cm}
\section{Conclusions}

Based on multiple observation-based evidences, we have argued that the complex radio emission in cluster Abell 980 arises mainly from two double radio sources created during two episodes of activity in the BCG, which has been drifting towards the cluster centre. Both lobes of the older radio source, spanning $\sim350$~kpc, are now diffuse. Their ultra-steep spectrum ($\alpha < -2$) and flattened appearance near the extremities are consistent with the notion that these two relic lobes have risen buoyantly towards the cluster periphery during their estimated spectral age of $\sim260$~Myr. The jet activity leading to the creation of these relic twin-lobes is probably also responsible for the two patches of enhanced ICM temperature observed between the lobes, consistent with recent numerical simulations. The axis of the younger double radio source coincident with the BCG is found to be parallel to the axis of the relic lobe pair, albeit at a lateral offset by $\sim70$~kpc. It is argued that the old (relic) lobe pair and the younger double radio sources are probably not independent sources, but, represent a double-double radio galaxy whose old and young lobe pairs have got laterally ‘detached’ due to the drift of their (common) parent galaxy towards the cluster centre. It is further noted that the two previously reported candidates for such detached Double-Double Radio Galaxy (dDDRG) seem to be less viable examples of this phenomenon.

\begin{acknowledgement}
G-K acknowledges a Senior Scientist fellowship from the Indian National Science Academy. SP would like to thank INSPIRE Faculty Scheme of DST, Govt. of India (code: IF-12/PH-44) for the research fund. S. Salunkhe would like to thank "Bhartratna JRD Tata Gunwant Sanshodhak Shishyavruti Yojna" for a doctoral fellowship. S. Sonkamble acknowledges the financial contribution from ASI-INAF n.2017-14-H.0 (PI A. Moretti). We thank Prof. Paul Wiita for his constructive comments on the manuscript.
\end{acknowledgement}



\appendix


\section{Radio spectral index error map}\label{apdx:radio-spectrum}

\begin{figure}
    \centering
    \includegraphics[width=0.95\textwidth]{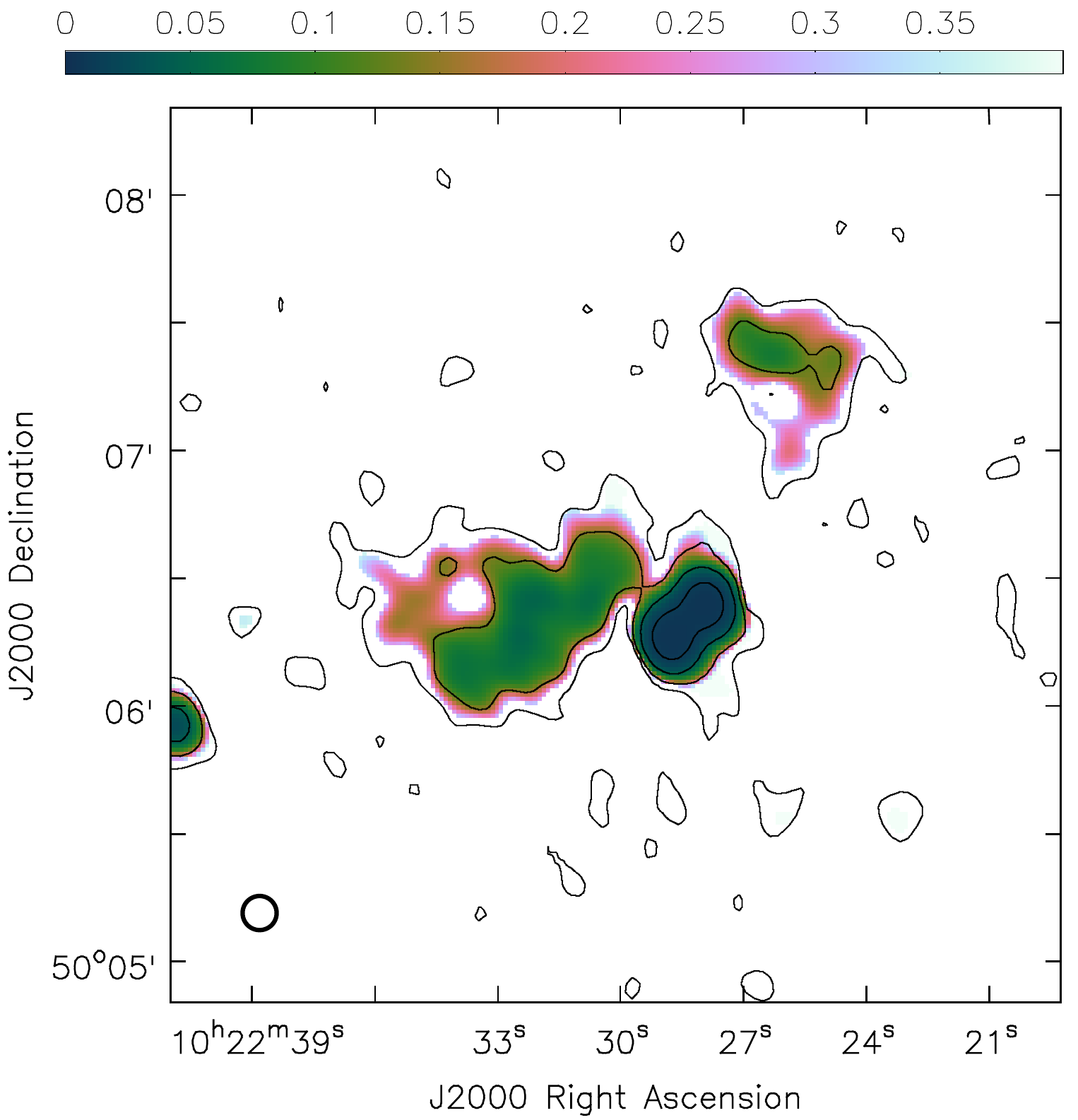}
    \caption{The error map for the spectral index map shown in Fig.~\ref{fig:Spectral_map_plot}a.}
    \label{fig:spix_err}
\end{figure}

\section{Spectral age error maps}\label{apdx:spectral_age_error}

\begin{figure*}
    \centering
    \includegraphics[width=0.48\textwidth]{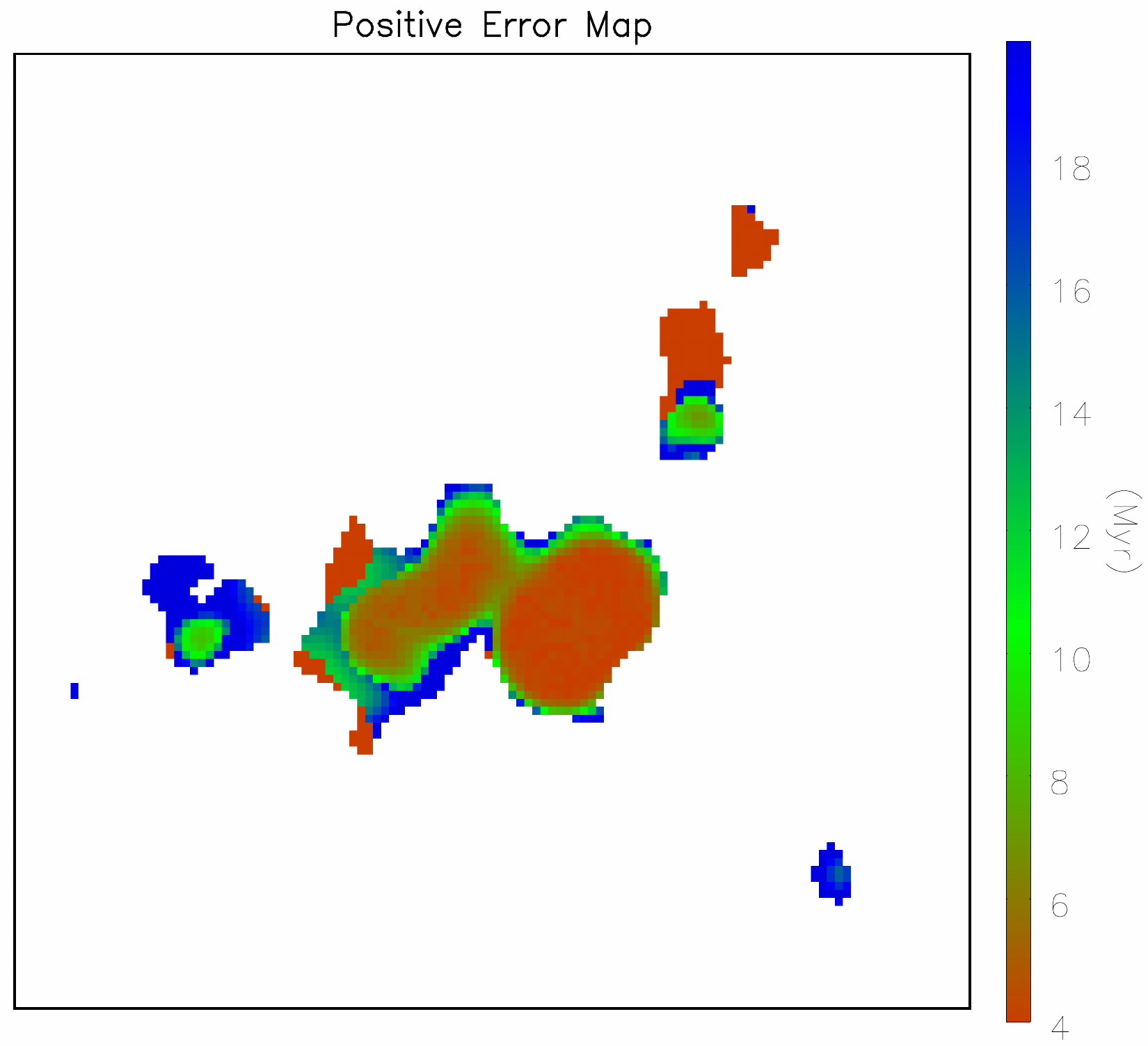}
     \includegraphics[width=0.48\textwidth]{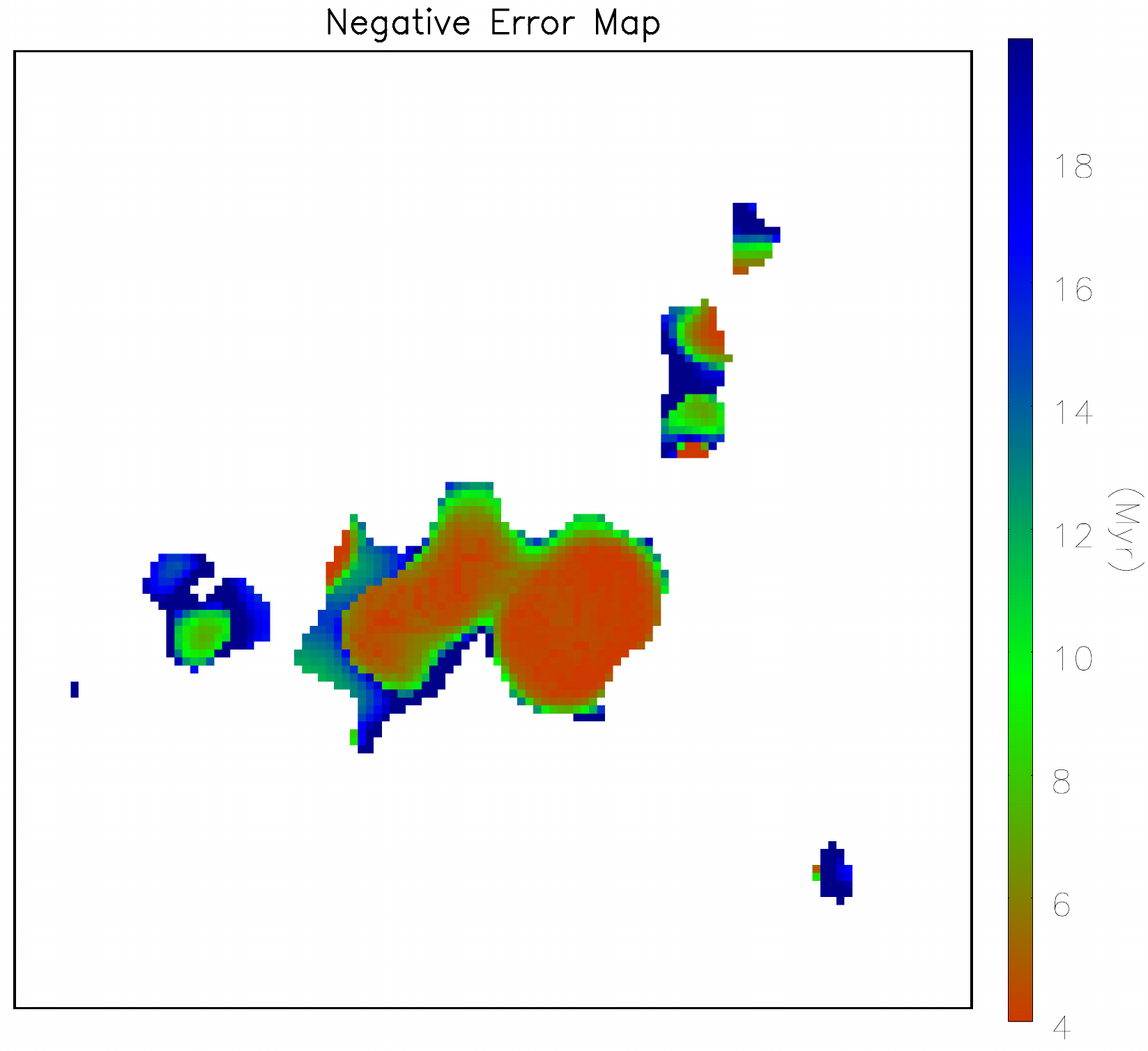}
    \caption{ Positive error map (left) and negative error map (right) for spectral age map presented in inset of Fig.~\ref{fig:Spectral_map_plot}a.}
    \label{fig:spix_err}
\end{figure*}

\end{document}